\documentclass[preprint,showpacs,preprintnumbers,amsmath,amssymb,aps]{revtex4}
\usepackage{graphicx}
\usepackage{dcolumn}
\usepackage{bm}
\usepackage[dvips]{color}
\begin{document}

\title{Magnetocrystalline anisotropy and magnetization reversal in $\mathbf{%
Ga}_{\mathbf{1-x}}\mathbf{Mn}_{\mathbf{x}}\mathbf{P}$ synthesized by ion
implantation and pulsed-laser melting}
\author{C. Bihler\footnote{bihler@wsi.tum.de}, M. Kraus, H. Huebl, and M. S. Brandt}
\affiliation{Walter Schottky Institut, Technische Universit\"{a}t M\"{u}nchen, Am
Coulombwall 3, 85748 Garching, Germany}
\author{S. T. B. Goennenwein and M. Opel}
\affiliation{Walther-Meissner-Institut, Bayerische Akademie der Wissenschaften,
Walther-Meissner-Str.~8, 85748 Garching, Germany}
\author{M. A. Scarpulla\footnote{present address: Materials Department, University of California, Santa
Barbara, CA 93106, mikes@engineering.ucsb.edu}, P. R. Stone, R. Farshchi, and
O. D. Dubon}
\affiliation{Department of Materials Science and Engineering, University of California,
Berkeley and Lawrence Berkeley National Laboratory, Berkeley, CA 94720}

\begin{abstract}
We report the observation of ferromagnetic resonance (FMR) and the
determination of the magnetocrystalline anisotropy in $\left(100\right)$%
-oriented single-crystalline thin film samples of $\mathrm{Ga}_{1-x}\mathrm{%
Mn}_x\mathrm{P}$ with $x=0.042$. The contributions to the magnetic
anisotropy were determined by measuring the angular- and the
temperature-dependencies of the FMR resonance fields and by superconducting
quantum interference device magnetometry. The largest contribution to the
anisotropy is a uniaxial component perpendicular to the film plane; however,
a negative contribution from cubic anisotropy is also found. Additional in-plane
uniaxial components are observed at low temperatures, which lift the
degeneracy between the in-plane $\left[011\right]$ and $\left[01\bar{1}%
\right]$ directions as well as between the in-plane $\left[010\right]$ and $\left[001%
\right]$ directions. Near $T=5$~K, the easy magnetization axis is close to
the in-plane $\left[01\bar{1}\right]$ direction. All anisotropy parameters
decrease with increasing temperature and disappear above the Curie
temperature $T_C$. A consistent picture of the magnetic anisotropy of
ferromagnetic $\mathrm{Ga}_{1-x}\mathrm{Mn}_x\mathrm{P}$ emerges from the
FMR and magnetometry data. The latter can be successfully modeled when both
coherent magnetization rotation and magnetic domain nucleation are considered.
\end{abstract}

\maketitle

\section{\label{sec:introduction}INTRODUCTION}

Mn-based diluted magnetic semiconductors show a variety of different
magnetic ordering phenomena ranging from ferromagnetism mediated by
quasi-delocalized holes in materials exhibiting metallic conductivity~\cite%
{Naz03, Edm04} to spin-glass like behavior in semiconducting matrices attributed to
Mn-rich nanoclusters~\cite{Jae06}. An important parameter expected to govern
the magnetic ordering is the localization of the charge carriers coupling
the 3$d$ high-spin states commonly introduced by Mn incorporation. In III-V
materials, where Mn simultaneously acts as acceptor, the corresponding
acceptor level essentially determines the degree of localization of the
holes~\cite{Gra03a}. A variation of the acceptor level can be achieved by
changing the group-V atom in III-V alloys. While recent studies on $\mathrm{%
Ga}_{1-x}\mathrm{Mn}_x\mathrm{N}$ indicate the formation of a ferromagnetic
ordering in essentially insulating material with a Curie temperature $T_C$
of 8~K~\cite{Sar06}, a carrier-mediated, non-metallic phase with $T_C$ up to 65 K in $\mathrm{Ga}%
_{1-x}\mathrm{Mn}_x\mathrm{P}$ has recently been synthesized~\cite{Sca05, Far06, Sto06a, Sto06}. In this material, it was
shown that $T_C$ increases with the magnetic dopant concentration and that
ferromagnetism is suppressed by the addition of compensating Te~\cite{Sca05} and S donors~\cite{Sto06a}. Furthermore, x-ray absorption spectroscopy and
x-ray magnetic circular dichroism have indicated that the local
ferromagnetic environment for Mn in $\mathrm{Ga}_{1-x}\mathrm{Mn}_x\mathrm{P}
$ is very similar to that in $\mathrm{Ga}_{1-x}\mathrm{Mn}_x\mathrm{As}$ and
that the Mn $d$-derived density of states at $E_F$ is strongly spin
polarized~\cite{Sto06}.

In $\mathrm{Ga}_{1-x}\mathrm{Mn}_x\mathrm{As}$, the magnetocrystalline
anisotropy has been successfully described in terms of the GaAs valence band
because the states occupied by holes mediating inter-Mn exchange appear to
be sufficiently similar to the unperturbed GaAs valence band~\cite{Die01}.
It has been previously pointed out \cite{Die00, Kep2004} that the holes
responsible for exchange mediation are probably at least semi-localized in
real space, as assumed in polaronic theories~\cite{Kam02, Dur02} also used
to describe diluted magnetic semiconductors. Evidence for some degree of
hole localization in III-Mn-V ferromagnetic semiconductors has been observed
in infrared studies of low-temperature molecular beam epitaxy (LT-MBE) grown 
$\mathrm{In}_{1-x}\mathrm{Mn}_x\mathrm{As}$~\cite{Hir01} and $\mathrm{Ga}%
_{1-x}\mathrm{Mn}_x\mathrm{As}$~\cite{Sin02}. In these materials it is
believed that there is significant mixing between the valence and Mn
impurity bands, while our studies in $\mathrm{Ga}_{1-x}\mathrm{Mn}_x\mathrm{P%
}$ suggest that the exchange-mediating holes reside in a separate Mn
impurity band~\cite{Sca05, Far06, Sto06a}. Conduction in this band occurs by
hopping in the ferromagnetic regime, indicating a much higher degree of
localization than in $\mathrm{Ga}_{1-x}\mathrm{Mn}_x\mathrm{As}$ or $\mathrm{%
In}_{1-x}\mathrm{Mn}_x\mathrm{As}$~\cite{Far06}. This suggests that, as predicted by ab-initio
calculations~\cite{Mah04, Sato2004, Jun2006, Kat07, Mas07}, a discontinuous transition to
an alternative mechanism of carrier-mediated exchange does not occur even
with these distinctions in band structure. As the magnetic anisotropy is
intimately tied to the properties of the exchange-mediating holes, it is important to investigate the magnetic anisotropy connected with the impurity band in $\mathrm{Ga}_{1-x}\mathrm{Mn}_x\mathrm{P}$.
Therefore, after giving a short introduction into the sample fabrication and
the measurement techniques in Sec.~\ref{sec:experimental}, we determine the
contributions to the magnetic anisotropy of $\mathrm{Ga}_{1-x}\mathrm{Mn}_x%
\mathrm{P}$ by measuring the angular- and the temperature-dependence of
ferromagnetic resonance (FMR) fields in Sec.~\ref{sec:FMR}. The FMR results
are substantiated by superconducting quantum interference device (SQUID)
magnetization measurements in Sec.~\ref{sec:SQUID}.

\section{\label{sec:experimental}EXPERIMENTAL}

Samples were prepared by ion implantation followed by
pulsed-laser melting (II-PLM)~\cite{Sca03a, Dub06}. For the present study, unintentionally sulfur-doped n-type GaP (100) wafers with a carrier
concentration in the range of $10^{16}~\mathrm{cm}^{-3}$ to $10^{17}~\mathrm{cm}^{-3}$ were
implanted with 50~keV Mn ions.
Each implanted sample was cleaved to have $\left[01\bar{1}\right]$ and $%
\left[011\right]$ edges and was irradiated in air with a single $0.4~\mathrm{%
J cm}^{-2}$ pulse ($\mathrm{FWHM}=23$~ns) from a KrF excimer laser ($\lambda
= 248$~nm) homogenized to a spatial uniformity of $\pm5\%$ by a
crossed-cylindrical lens homogenizer. Channeling $^{4}\mathrm{He}^+$
Rutherford backscattering spectrometry (RBS) and particle induced x-ray
emission (PIXE) were used to assess the crystalline quality, Mn dose retained after II-PLM,
and substitutional fraction of Mn in the samples~\cite{Dub06}. Once processed the films are high-quality single crystals with a Mn dose of $7.3 \times 10^{15}$~cm$^{-2}$ and substitutionality of $0.7$, i.e., $70\%$ of Mn atoms substitute Ga atoms.  We note that the remaing $30\%$ of Mn atoms do not form interstitial defects and instead are incommensurate with the GaAs lattice presumably in the form of small clusters.  This level of substitutionality is not unlike $\mathrm{Ga}_{1-x}\mathrm{Mn}_x\mathrm{As}$ films of higher Mn concentration grown by low-temperature molecular beam epitaxy, which contain on the order of $20\%$ non-substitutional Mn~\cite{Yu02}.

II-PLM processing results in samples having a gradient in Mn concentration into the depth
of the sample as measured by secondary ion mass spectrometry (SIMS), making
it impossible to determine single values for the film thickness and Mn
concentration. The Mn SIMS profile can be approximated by a Gaussian
distribution centered at a depth of 40~nm with a width of 20~nm. However, as
the regions of the film with highest Mn concentration dominate both the
magnetic and transport properties, samples are discussed here in terms of
their peak \textit{substitutional} Mn concentration, which was determined by
channeling RBS and PIXE to be $x=0.042$~\cite{Sca05, Far06}. Sample magnetization was determined in various crystallographic
orientations using a superconducting quantum interference device (SQUID)
magnetometer. The FMR measurements were performed at $\omega/2\pi\approx9.3~%
\mathrm{GHz}$ in an electron paramagnetic resonance (EPR) spectrometer using
magnetic field modulation, with the sample temperature controlled using a
liquid-He flow cryostat.

\section{\label{sec:FMR}FERROMAGNETIC RESONANCE SPECTROSCOPY}

\begin{figure}[tbp]
\includegraphics[width=0.7\textwidth]{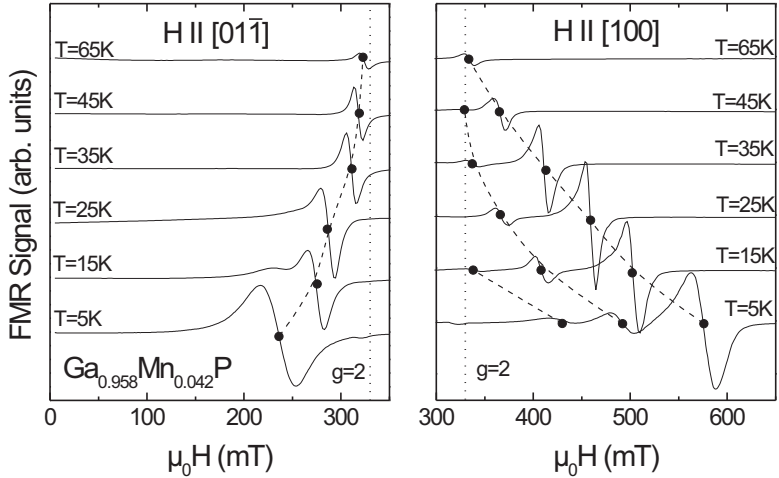} 
\caption{Temperature dependence of the FMR signal of a $\mathrm{Ga}_{0.958}%
\mathrm{Mn}_{0.042}\mathrm{P}$ sample for the magnetic field aligned along
the in-plane $[01\bar{1}]$ (left panel) and the out-of-plane $[100]$ (right
panel) directions. The dashed lines are guides to the eye. The magnetic
field corresponding to $g=2$ is indicated by dotted vertical lines.}
\label{fig:FMRvT}
\end{figure}

\begin{figure}[tbp]
\includegraphics[width=0.7\textwidth]{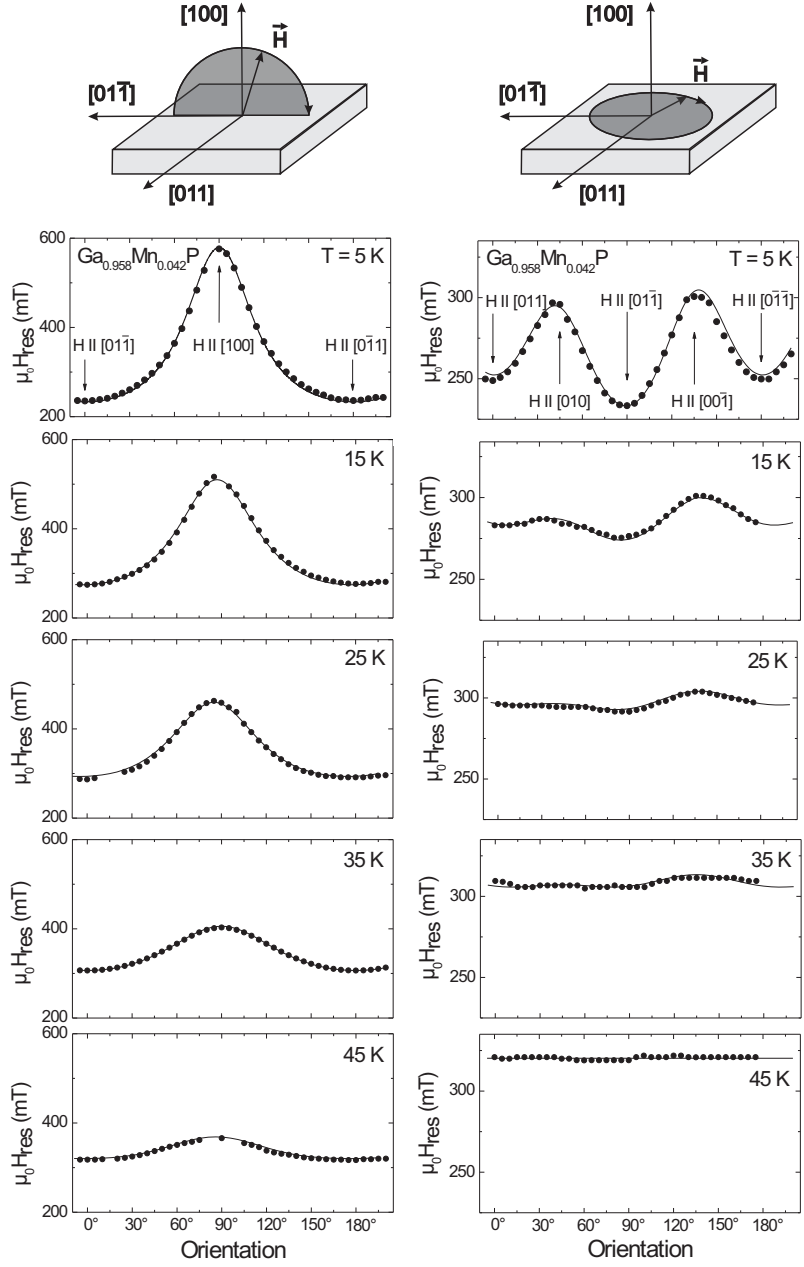} \vspace{0.5cm}
\caption{Angular dependence of the ferromagnetic resonance fields for the
magnetic field rotating within the $\left(011\right)$ (left panels) and $%
\left(100\right)$ (right panels) planes at $T=5$~K, 15~K, 25~K, 35~K, and
45~K. The circles correspond to the experimentally observed resonance
positions. The full lines show the anisotropy expected for the parameters
shown in Fig.~\protect\ref{fig:Aniso}.}
\label{fig:Stern}
\end{figure}

The left panel of Fig.~\ref{fig:FMRvT} shows the temperature dependence of
the FMR signal of a typical $x=0.042$ sample for the magnetic field aligned
along the in-plane $[01\bar{1}]$ direction, while the right panel presents
data from the out-of-plane $\left[100\right]$ direction. At $T=5$~K for $H$
parallel to the in-plane $[01\bar{1}]$ direction, we observe one resonance
at $\mu_0H=236$~mT with a peak-to-peak linewidth of $\mu_0\Delta
H_{pp}\approx36$~mT, while for $H$ perpendicular to the sample surface ($%
H~||~[100]$) there are three distinct resonances at $\mu_0H=576$~mT, $492$%
~mT, and $430$~mT, each with $\mu_0\Delta H_{pp}\approx25$~mT. With
increasing temperature, the resonance fields for both orientations shift
toward $\mu_0H=330$~mT, which corresponds to the resonance field of
paramagnetic impurities with a $g$-factor of $g=2$. The anisotropy
disappears around $T\approx65$~K, slightly above the Curie temperature $T_C=55$~K determined from SQUID magnetization measurements. We attribute this to the moderately-large applied field $\mu_0H\approx330$~mT in resonance, which stabilizes ferromagnetism even
slightly above $T_C$ (compare Fig.~3 in Ref.~\cite{Sto06}).

We attribute the multiple resonances in the $H~||~[100]$ data to spin wave
excitations~\cite{Gön03b, Kit58, Koe03,Rap04}. Since we only observe three resonances or less, a
detailed analysis of the mode spacing is hardly possible. However, assuming
the resonance at the highest magnetic field $\mu_0H_0=576$~mT to be the
collective mode, we determine a separation $\Sigma(n)=H_0-H_n\propto n^{0.8}$
between the resonance fields of the mode with the highest field $H_0$ and
the $n$th spin wave mode $H_n$ at $T=5$~K, which does not obey the classical
behavior expected for a homogeneous film $\Sigma_{\mathrm{classical}%
}(n)\propto n^2$~\cite{Kit58}. We attribute this to the varying depth
profile of the Mn concentration. A similar non-quadratic behavior of the
mode spacing has been reported for $\mathrm{Ga}_{1-x}\mathrm{Mn}_x\mathrm{As}
$ thin films exhibiting gradients in hole concentration \cite{Gön03b, Koe03,
Rap04}. Describing the implantation profile with a parabolic depth
dependence of the Mn concentration, a linear modes spacing would be expected,
in reasonable agreement with the observed behavior.

\begin{figure}[tbp]
\includegraphics[width=0.3\textwidth]{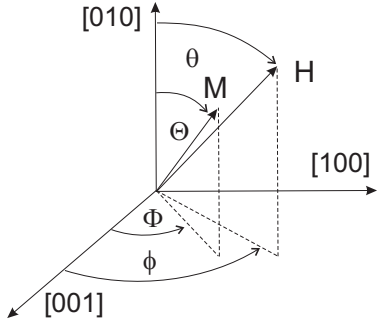}
\caption{Coordinate system used for FMR simulation, with the orientation of
the saturation magnetization $M=M(\Theta,\Phi)$ and the magnetic field $H=H(%
\protect\theta,\protect\phi)$.}
\label{fig:Koord}
\end{figure}

To elucidate the magnetic anisotropy of the samples we performed
measurements of the angular dependence of the FMR for sample rotations in
the $\left(011\right)$ and $\left(100\right)$ planes. The angular
dependences of the resonance fields obtained at different temperatures are
shown in Fig.~\ref{fig:Stern}. The panels on the left hand side correspond
to rotations of the external magnetic field from the in-plane $\left[01\bar{1%
}\right]$ to the out-of-plane $\left[100\right]$, and back to the in-plane $%
\left[0\bar{1}1\right]$ direction. For simplicity, we limit the discussion
to the collective mode in the following. A uniaxial magnetic anisotropy with the
magnetic hard axis perpendicular to the layer can be inferred from the
increase in ferromagnetic resonance fields approaching $[100]$. Likewise,
the fourfold symmetry observed for the in-plane rotations in the right hand
panels demonstrates the presence of a cubic anisotropy contribution.
Additionally, the fact that the $\left[01\bar{1}\right]$ and $\left[011%
\right]$ orientations, as well as the $\left[010\right]$ and $\left[001%
\right]$ orientations are not degenerate indicates the contribution of
further in-plane uniaxial anisotropy components.

For a quantitative simulation of these data we use the free energy density 
\begin{eqnarray}  \label{FreeE}
F =&-&M H(\sin\Theta \sin\Phi \sin\theta \sin\phi + \cos\Theta \cos\theta +
\sin\Theta \cos\Phi \sin\theta\cos\phi)  \nonumber \\
&+& K^{100}_{\mathrm{eff}}\sin^2\Theta \sin^2\Phi  \nonumber \\
&-& \frac{1}{2} K^{\bot}_{\mathrm{c}1} \sin^4\Theta \sin^4\Phi  \nonumber \\
&-& \frac{1}{2} K^{||}_{\mathrm{c}1} (\cos^4\Theta + \sin^4\Theta\cos^4\Phi) 
\nonumber \\
&+& \frac{1}{2} K^{011}_{\mathrm{u}} (\cos\Theta +\sin\Theta \cos\Phi)^2  \nonumber \\
&+& K^{001}_{\mathrm{u}} \sin^2\Theta \cos^2\Phi.
\end{eqnarray}
The angles are given by the orientation of the saturation magnetization $%
M=M(\Theta,\Phi)$ and the magnetic field $H=H(\theta,\phi)$ (Fig.~\ref%
{fig:Koord}). The first term describes the Zeeman energy, while the second
term represents an effective perpendicular uniaxial anisotropy $%
K^{100}_{\mathrm{eff}} $ and is composed of the sum of demagnetization and uniaxial
magnetocrystalline components, $\frac{1}{2}\mu_0M^2$ and $K^{100}_{\mathrm{u}}$,
respectively. In order to describe the breaking of the cubic symmetry due to
in-plane biaxial compressive strain induced by the presence of Mn, we
include separate cubic terms for in-plane and perpendicular components,
given by $K^{||}_{\mathrm{c}1}$ and $K^{\bot}_{\mathrm{c}1}$, respectively. The in-plane cubic
symmetry breaking is accounted for by the final two terms representing
uniaxial contributions along $\left[011\right]$ ($K^{011}_{\mathrm{u}}$) and $\left[001%
\right]$ ($K^{001}_{\mathrm{u}}$). In all cases the appropriate anisotropy field is
given by the ratio $2K/M$. The come about of the different anisotropy terms, as well as of the equivalence of first order cubic and second order uniaxial anisotropy, are discussed in Appendix A.

\begin{figure}[tbp]
\includegraphics[width=0.7\textwidth]{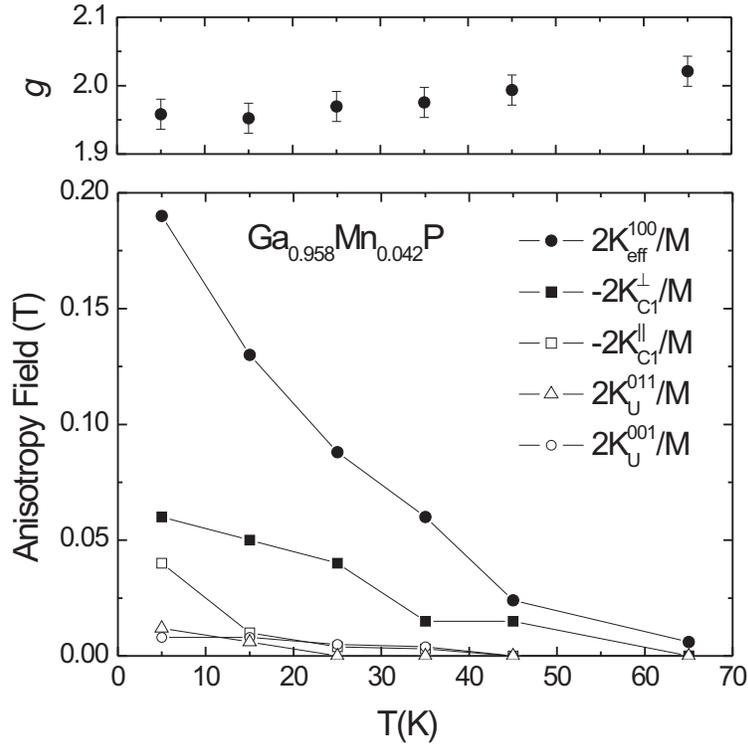} 
\caption{Temperature dependence of the anisotropy parameters and the $g$%
-factor obtained from the simulation of the angular dependence of the
ferromagnetic resonance (full lines in Fig.~\protect\ref{fig:Stern}).}
\label{fig:Aniso}
\end{figure}

Following the approach of Smit \textit{et al.,}\cite{Smi54, Smi55} we obtain
the equation of motion 
\begin{eqnarray}
\left(\frac{\omega}{\gamma}\right)^2=\frac{1}{M^2\sin^2\Theta}\left[\left(%
\frac{\partial^{2}}{\partial\Phi^{2}}F\right)\left(\frac{\partial^{2}}{%
\partial\Theta^{2}}F\right)\right.\left.-\left(\frac{\partial}{\partial\Phi}%
\frac{\partial}{\partial\Theta}F\right)^2\right]_{\Phi_0,\Theta_0}
\end{eqnarray}
with the gyromagnetic ratio $\gamma=\frac{g\mu_B}{\hbar}$, which has to be
evaluated at the equilibrium orientation of the saturation magnetization
determined from 
\begin{eqnarray}
\frac{\partial}{\partial\Phi} F |_{ \Phi=\Phi_0 } =\frac{\partial}{%
\partial\Theta} F |_{ \Theta=\Theta_0 }=0.
\end{eqnarray}
The solution of these equations yields the FMR resonance condition. The full
lines in Fig.~\ref{fig:Stern} are simulations of the measured data with
anisotropy fields plotted in Fig.~\ref{fig:Aniso}. At 5 K, the results for
four samples with $x=0.042$ are well reproducible, with $0.16~\mathrm{T}%
<2K^{100}_{\mathrm{eff}}/M<0.19~\mathrm{T}$, $-0.10~\mathrm{T}<2K^{\bot}_{\mathrm{c}1}/M<-0.06~%
\mathrm{T}$, $-0.04~\mathrm{T}<2K^{||}_{\mathrm{c}1}/M<-0.03~\mathrm{T}$, $0~\mathrm{T%
}<2K^{011}_{\mathrm{u}}/M<0.012~\mathrm{T}$, and $0~\mathrm{T}<2K^{001}_{\mathrm{u}}/M<0.02~%
\mathrm{T}$. The magnetic anisotropy of these films is clearly dominated by
the uniaxial and cubic contributions perpendicular to the film. The in-plane
cubic and uniaxial anisotropies along $[011]$ and $[001]$ are all somewhat
smaller. Figure~\ref{fig:Aniso} depicts the decrease of all of the
anisotropy components with increasing temperature and demonstrates that all
components disappear above the $T_C$ of 55~K of the films as expected. As discussed above, only a
small $2K^{100}_{\mathrm{eff}}$ is observed even above $T_C$ at 65~K.

From SQUID magnetization measurements we estimate the saturation
magnetization in the most heavily doped part of the film $-$ i. e. near the
peak of the Mn distribution $-$ to $M=37$~kA/m. From this, we obtain an
upper limit for the demagnetization field $\mu _{0}M$ of 0.05~T. Therefore, the demagnetization field constitutes only about one fourth of the effective uniaxial magnetic anisotropy field along $[100]$, $2K_{\mathrm{eff}}^{100}/M = 0.19$~T, determined from the simulation of the angular dependence of FMR. This strongly indicates the presence of a tetragonal distortion of the $%
\mathrm{Ga}_{0.958}\mathrm{Mn}_{0.042}\mathrm{P}$ layer after pulsed-laser
melting causing the dominating contribution $K_{\rm u}^{100}$ as in the case of $%
\mathrm{Ga}_{1-x}\mathrm{Mn}_{x}\mathrm{As}$, where a strong uniaxial
magnetic anisotropy in the growth direction is commonly observed and is
attributed to the tetragonal distortion of the $\mathrm{Ga}_{1-x}\mathrm{Mn}%
_{x}\mathrm{As} $ layer due to lattice-matched growth~\cite{Liu03,
Bihler3112006, Lim06}. However, a quantitative detection of the distortion in case of $%
\mathrm{Ga}_{0.958}\mathrm{Mn}_{0.042}\mathrm{P}$ via x-ray diffraction
turns out to be difficult due to the inhomogeneous Mn profile after pulsed-laser melting and the accompanied broadening of the diffraction peaks.

One very interesting finding is that the magnetic easy axes of the in-plane
cubic magnetic anisotropy are along $\left[ 01\bar{1}\right] $ and $\left[
011\right] $ directions as opposed to the $\left[ 010\right] $ and $\left[
001\right] $ directions commonly observed for LT-MBE grown $\mathrm{Ga}_{1-x}%
\mathrm{Mn}_{x}\mathrm{As}$. This observation gives rise to the negative
sign of $K_{\mathrm{c}1}^{||}$ in the $\mathrm{Ga}_{0.958}\mathrm{Mn}_{0.042}\mathrm{P%
}$ samples studied ($K_{\mathrm{c}1}^{\bot }$ is also negative). To the best of our
knowledge, a negative cubic anisotropy has so far only been reported for $%
\mathrm{In}_{1-x}\mathrm{Mn}_{x}\mathrm{As}$~\cite{Liu05InMnAs, Saw04}.
Within the different models for carrier-mediated ferromagnetism in diluted
magnetic semiconductors~\cite{Die01, Abo01}, the sign of the cubic
anisotropy is predicted to oscillate with varying hole concentration.
However, the applicability of these models for the strongly localized,
impurity-band-like character expected for the holes in $\mathrm{Ga}_{1-x}%
\mathrm{Mn}_{x}\mathrm{P}$ remains an open question. Taking into account all
anisotropy contributions, the global magnetic easy axis at 5~K is oriented
close to the $\left[ 01\bar{1}\right] $ direction.

Interestingly, the $g$-factor does not deviate significantly from $g=2$. In $%
\mathrm{Ga}_{1-x}\mathrm{Mn}_{x}\mathrm{As}$, $g$ was found to be an
effective $g$-factor taking into account both the contributions of the Mn
atoms and the hole subsystem~\cite{Liu05}. Depending on the hole
concentration $p$, Liu \textit{et al.} observed a $g$-factor at 4.2~K between $%
g=1.80$ for a sample with $p=1.64\times 10^{20}$~cm$^{-3}$ and $g=1.95$ for
a sample with $p=1.24\times 10^{20}$~cm$^{-3}$~\cite{Liu05}. The fact that $%
g $ is found to vary from 1.95 to 2 for increasing temperature from 5~K to
65~K indicates that there is only an even smaller contribution of the hole
subsystem to the effective $g$-factor in the case of $\mathrm{Ga}_{1-x}%
\mathrm{Mn}_{x}\mathrm{P}$. This is consistent with the observations and
calculations in Ref.~\cite{Far06} indicating a small hole concentration of
up to $10^{20}$cm$^{-3}$.

\section{\label{sec:SQUID}SQUID MAGNETIZATION MEASUREMENTS}

\begin{figure}[tbp]
\includegraphics[width=0.6\textwidth]{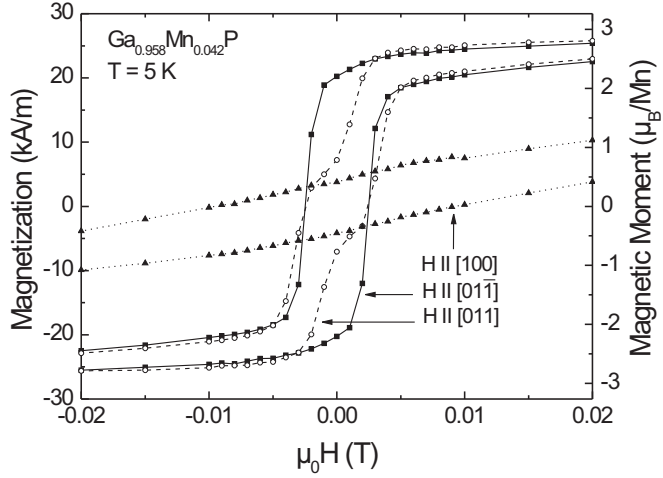} 
\caption{$M(H)$ SQUID magnetization curves along the out-of-plane $\left[100%
\right]$ (solid triangles) and the in-plane $\left[01\bar{1}\right]$ (solid
squares) and $\left[011\right]$ (open circles) crystallographic axes. The
lines are guides to the eye. The right and left vertical axes give the magnetic moment per \textit{substitutional} Mn atom $m_{\rm Mn}$ and the magnetization $M$ of the sample in the region of highest Mn concentration, respectively. These values were deduced from the measured total magnetic moment $m_{\rm tot}$ as described in Appendix B.}
\label{fig:hysges1}
\end{figure}

To substantiate the results obtained from FMR in the preceding section we
also performed SQUID magnetization measurements. Figure~\ref{fig:hysges1}
compares the $M(H)$ magnetization curves obtained at $T=5$~K for the
external magnetic field oriented along the out-of-plane $\left[ 100\right] $
(solid triangles) and the in-plane $\left[ 01\bar{1}\right] $ (solid
squares) and $\left[ 011\right] $ (open circles) crystallographic axes. The
square-like $M(H)$ curve obtained for $H||\left[ 01\bar{1}\right] $ also
indicates that $\left[ 01\bar{1}\right] $ is the easy magnetic axis at 5~K.
Similarly, the large field of $\approx 0.2$~T required to align $M$ along
the magnetically hard $\left[ 100\right] $ direction is due to the large
out-of-plane uniaxial and cubic anisotropy contributions. In the following,
we use the free energy ansatz of (\ref{FreeE}) to simulate these $M(H)$
curves and especially to explain the kink observed for the magnetization
measurement along $\left[ 011\right] $.

\begin{figure}[tbp]
\includegraphics[width=0.5\textwidth]{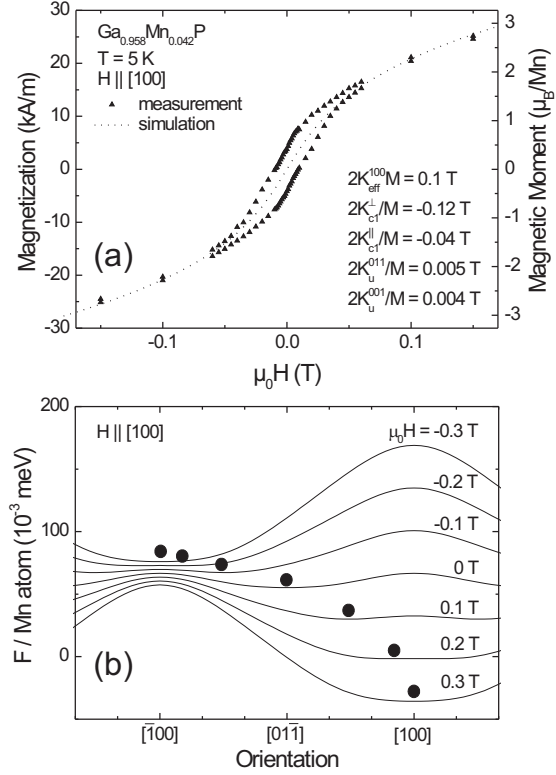} 
\caption{(a) Comparison of the $M(H)$ SQUID magnetization curve measured for 
$H$ along the out-of-plane $\left[100\right]$ direction (full triangles)
with the curve simulated using the anisotropy parameters given in the figure
(dotted line). (b) Free energy per Mn atom as a function of the orientation
of magnetization in the (011) plane for different magnetic fields applied
along $\left[100\right]$ calculated from (\ref{FreeE}). The position of the solid circle corresponds to
the equilibrium orientation of magnetization in the minimum of the free
energy surface. For clarity, the curves are shifted vertically.}
\label{fig:oop2}
\end{figure}

To begin with, we focus on the $M(H)$ curve for $H||\left[ 100\right] $ (out
of plane) shown in Fig.~\ref{fig:oop2}(a) on a larger field scale. The
dotted line is the curve simulated as discussed below for which we obtained
the best agreement with the SQUID measurement using the anisotropy fields $%
2K_{\mathrm{eff}}^{100}/M=0.1$~T, $2K_{\mathrm{c}1}^{\bot }/M=-0.12$~T, $2K_{\mathrm{c}1}^{||}/M=-0.04$%
~T, $2K_{u}^{011}/M=0.005$~T, and $2K_{u}^{001}/M=0.004$~T in (\ref{FreeE}).
For $H||\left[ 100\right] $ the simulated curve is predominantly determined
by $2K_{\mathrm{eff}}^{100}/M$ and $2K_{\mathrm{c}1}^{\bot }/M$. Both these parameters agree
with the ones determined from FMR, to within a factor of two which can be
understood as follows. The FMR spectra for $H$ oriented along the hard
magnetic out-of-plane $\left[ 100\right] $ axis features spin wave
excitations. This not only indicates that there is a gradient in magnetic
properties as already discussed above, but also that these modes are located
at the region of the highest uniaxial anisotropy field $2K_{\mathrm{eff}}^{100}/M$
and therefore only probe the magnetic properties of this specific region.
SQUID magnetization measurements in contrast integrate over the magnetic
properties of the whole layer. In this respect the agreement of FMR and
SQUID within a factor of two is quite reasonable.

The curvature of the simulated magnetization curve can be understood looking
at the dependence of the free energy per Mn atom on the orientation of the
magnetization in the (011) plane for different magnetic fields applied along 
$\left[100\right]$ [Fig.~\ref{fig:oop2}(b)]. The position of the solid
circle corresponds to the equilibrium orientation of magnetization in the
minimum of the free energy surface. For high magnetic fields $H || \left[100%
\right]$, the magnetization is also in the $\left[100\right]$ direction,
since then the Zeeman term is the dominant contribution in (\ref{FreeE}).
This is the case for $\mu_0H=0.3$~T in Fig.~\ref{fig:oop2}(b). For
decreasing magnetic fields the magnetocrystalline anisotropy becomes increasingly important. This leads to the migration of the minimum in the free
energy surface $-$ and therefore also equilibrium orientation of the
magnetization $-$ in the direction of the $\left[01\bar{1}\right]$ axis,
which is the magnetic easy axis for zero magnetic field. The application of
a magnetic field in the opposite direction (i.e. $H || \left[\bar{1}00\right]
$) in turn tilts the magnetization more and more in this direction [Fig.~\ref%
{fig:oop2}(b)]. Having determined the orientation of the magnetization
depending on the field strength $H$, we obtain the simulated curve in Fig.~%
\ref{fig:oop2}(a) via calculating the projection of the magnetization along
the direction of the external magnetic field, which is the quantity measured
by the SQUID magnetometer. The process of magnetization reversal described
in this paragraph is called \textit{coherent spin rotation}~\cite{Liu2005}.

\begin{figure}[tbp]
\includegraphics[width=0.5\textwidth]{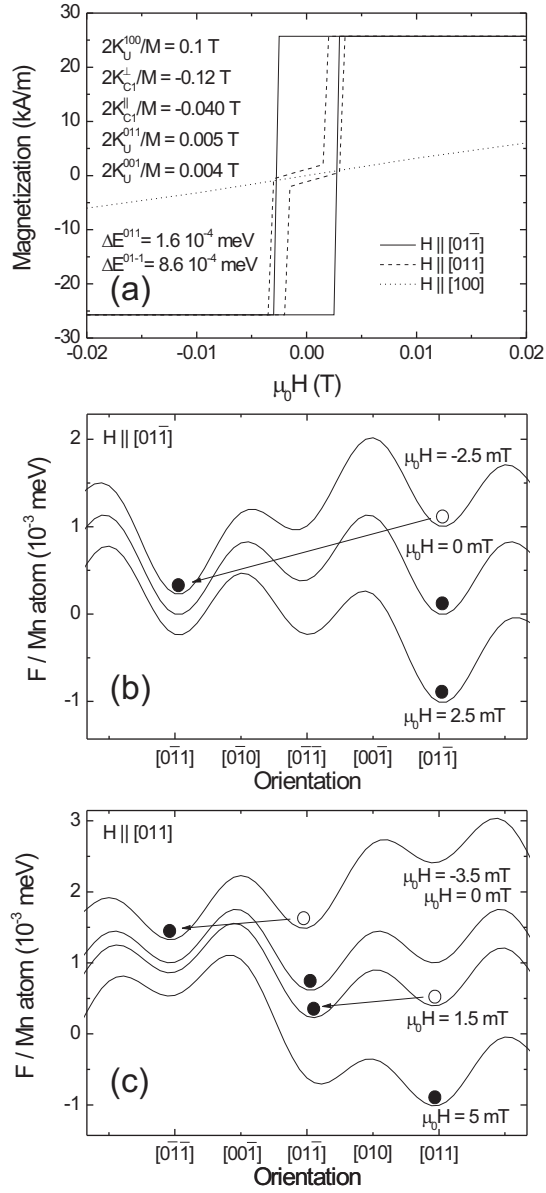} 
\caption{(a) Simulated magnetization curves using the anisotropy parameters
given in the figure. The solid, dashed, and dotted curves correspond to $H
|| \left[01\bar{1}\right]$, $H || \left[011\right]$, and $H || \left[100%
\right]$, respectively. (b) Free energy per Mn atom as a function of the
orientation of magnetization in the (100) film plane for different magnetic
fields applied along $\left[01\bar{1}\right]$ and (c) along $\left[011\right]
$. The position of the solid circle corresponds to the equilibrium
orientation of magnetization in the minimum of the free energy surface. For
clarity the curves are shifted vertically.}
\label{fig:ip2}
\end{figure}

For the simulation of the in-plane $M(H)$ curves in Fig.~\ref{fig:ip2}(a),
in addition to coherent spin rotation, the process of \textit{non-coherent
spin switching} has to be considered following the discussion of Ref.~\cite%
{Liu2005}. For $H||\left[ 01\bar{1}\right] $, spin switching is visualized
in Fig.~\ref{fig:ip2}(b), where the dependence of free energy per Mn atom on
the orientation of the magnetization in the (100) film plane is plotted for
different magnetic fields. Decreasing the external magnetic field $H||\left[
01\bar{1}\right] $ from its maximum value of 7~T to zero, the magnetization
remains "trapped" in the global minimum at $\left[ 01\bar{1}\right] $. For
negative fields, $\left[ 01\bar{1}\right] $ turns into a local minimum,
while $\left[ 0\bar{1}1\right] $ becomes the global minimum in free energy.
Since the thermal energy $k_{B}T$ at $T=5$~K of $0.43$~meV is three orders
of magnitude larger than the energy needed to overcome the barrier between
the two minima, there will be always some magnetic moments oriented along
the direction of the global minimum. However, for the generation of a new
magnetic domain with a magnetization along $\left[ 0\bar{1}1\right] $, the
domain walls of the nucleus of this domain first have to be formed. The
formation of these domain walls is energetically unfavorable, since the
magnetic moments in the walls are oriented along the magnetic harder axes of
the energy barriers. Therefore, magnetization reversal takes place only if
the energy gain from tilting the magnetization into the direction of the
global minimum accounts for the energy needed for the formation the domain
walls of this new magnetic domain. To obtain the experimentally observed
switching field of -2.5~mT found for $M(H)$, with $H||\left[ 01\bar{1}\right]
$, we have to assume a domain wall formation energy of $\Delta E^{01\bar{1}%
}=8.6\times 10^{-4}$~meV per Mn atom.

\begin{figure}[tbp]
\includegraphics[width=0.6\textwidth]{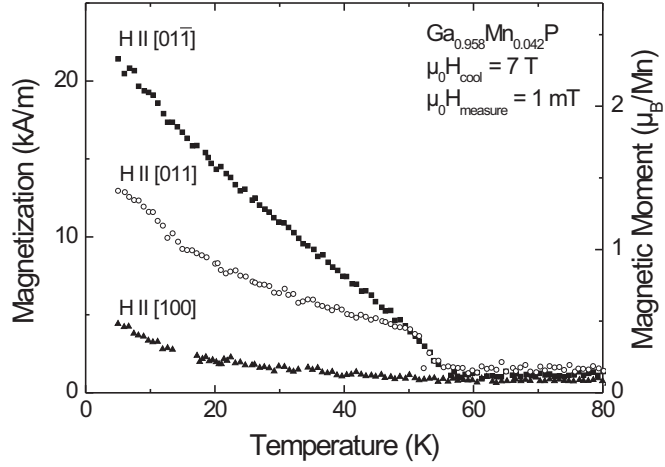} 
\caption{Temperature dependence of magnetization for the sample cooled down
to 5~K in a field of $\protect\mu_0H_{cool}=7$~T and measured during warm up
in a field of $\protect\mu_0H_{measure}=1$~mT along the same
crystallographic axis as in Fig.~\protect\ref{fig:hysges1}.}
\label{fig:hysges2}
\end{figure}

The situation for $H||\left[ 011\right] $ is plotted in Fig.~\ref{fig:ip2}%
(c). Decreasing the external magnetic field $H||\left[ 011\right] $ from its
maximum value of 7~T to zero, the magnetization first also remains "trapped"
in the global minimum at $\left[ 011\right] $. However, approaching $\mu
_{0}H=0$~mT, $\left[ 01\bar{1}\right] $ becomes the global minimum due to
the presence of the uniaxial anisotropy field along $\left[ 011\right] $.
Therefore, there will be a first switching into the $\left[ 01\bar{1}\right] 
$ direction at positive fields and a second switching into the $\left[ 0\bar{%
1}\bar{1}\right] $ direction at large enough negative magnetic fields. In
the model, we assumed the same energy barrier $\Delta E^{011}$ for both
switching processes. Then, the parameters predominantly determining the
switching fields are $\Delta E^{011}$ and $2K_{u}^{011}/M$. For increasing $%
\Delta E^{011}$ both switching processes occur later (at lower fields),
while for increasing $2K_{u}^{011}/M$ the first switching process takes
place earlier (at higher field) and the second one later (at lower field).
The best agreement with the experimentally observed switching fields, $\mu
_{0}H=1.5$~mT for the first and $\mu _{0}H=-3.5$~mT for the second switching
[compare Fig.~\ref{fig:hysges2}], we obtained for the domain wall formation
energy of $\Delta E^{011}=1.6\times 10^{-4}$~meV per Mn atom and the
uniaxial anisotropy field $2K_{u}^{011}/M=5$~mT. Note that in contrast to
the 180$^{\circ }$ domain walls that must be formed in case of the
magnetization reversal for $H||\left[ 01\bar{1}\right] $, the domain walls
that must be formed here comprise a 90$^{\circ }$ rotation of magnetization.
Therefore it is not surprising that $\Delta E^{011}<\Delta E^{01\bar{1}}$.
Typical values we obtain for the height of the energy barrier separating two
local energy minima $\Delta F=7\times 10^{-4}$~meV per Mn ion are one order
of magnitude smaller than the value obtained for a perpendicular
magnetization-reversal process in case of $\mathrm{Ga}_{1-x}\mathrm{Mn}_{x}%
\mathrm{As}$ by Liu \textit{et al.}~\cite{Liu2005}. Applying Kittel's Bloch
domain wall model as discussed by Liu \textit{et al.}~\cite{Liu2005} to our in-plane
magnetization reversal process for $H||\left[ 011\right] $ we obtain a size
of a domain nucleus of approximately $5~\mu $m. However, note that the
assumption of a Bloch domain wall may not be justified in this case of an
in-plane magnetized ferromagnetic film.

Consequently, the kink in the $M(H)$ curve for $H || \left[011\right]$ can
be explained by the fact that due to the presence of the uniaxial anisotropy
field along $\left[011\right]$, $\left[011\right]$ is not the global easy
magnetic axis. At low fields a non-coherent spin switching into the global
easy magnetic axis along $\left[01\bar{1}\right]$ takes place, which causes
a vanishing projection of $M$ on the $\left[011\right]$ direction. The fact
that there still is a finite projection in Fig.~\ref{fig:ip2}(a) is caused
by the presence of the uniaxial in-plane anisotropy field $2K^{001}_{\rm u}/M$ which slightly
changes the 90$^{\circ}$ angle between the two minima of the cubic anisotropy
near $\left[01\bar{1}\right]$ and $\left[011\right]$.

It should be noted that our simulation only accounts for hysteresis effects
caused by the non-coherent spin switching described above; the additional
hysteresis effects observed for example for the $M(H)$ curve measured for H
along the out-of-plane [100] direction are presumably caused by the pinning
and de-pinning of domain walls at crystal defects, which is not included in our model.
Furthermore, the magnetization is by far not saturated at magnetic fields of 
$\mu _{0}H=0.02$~T. Therefore, the saturation magnetization used in our model $M=26$~kA/m is lower than the real saturation magnetization of $M=37$~kA/m at $\mu _{0}H=7$~T. Internal stresses, or shape irregularities could
explain the rounding of the magnetization curve at high fields~\cite{Hub98}. However, in
spite of the simplicity of the model, the $M(H)$ magnetization curves along
several crystallographic directions can be explained at least semi-quantitatively by the
presence of the anisotropy fields determined from FMR.

Finally, we discuss the temperature dependence of magnetization along
different crystallographic orientations (Fig.~\ref{fig:hysges2}). In these
measurements the sample is cooled down in the maximum available field $\mu
_{0}H_{\rm cool}=7$~T. At 5~K, the field is switched to $\mu _{0}H_{\rm measure}=1$%
~mT and the projection of magnetization along the field direction is
measured during warm up of the sample. For $H||\left[ 01\bar{1}\right] $
(closed squares), we obtained the highest value for this projection at all
temperatures in agreement with $\left[ 01\bar{1}\right] $ being the easy
magnetic axis. Accordingly, the projection along $\left[ 100\right] $
(closed triangles) is very small, since $\left[ 100\right] $ is the hard
magnetic axis. For $H||\left[ 011\right] $ (open circles) and $T<50$~K, the
projection lies in between the values for the preceding orientations. This
can be explained by the fact that $\left[ 011\right] $ is not the global
easy magnetic axis in this temperature range. Above 50~K, the curve for $H||%
\left[ 011\right] $ overlaps with the one for $H||\left[ 01\bar{1}\right] $,
which is in good agreement with the disappearance of the uniaxial anisotropy
field $2K_{u}^{011}/M$ in this temperature range (see Fig.~\ref{fig:Aniso}),
which lifts the degeneracy between the easy axes $\left[ 01\bar{1}\right] $
and $\left[ 011\right] $. Furthermore, the Curie temperature $T_{C}=55$~K
deduced from the $M(T)$ curves in Fig.~\ref{fig:hysges2} again nicely agrees
with the temperature around which the anisotropy fields vanish in Fig.~\ref%
{fig:Aniso}.

\section{\label{sec:conclusion}CONCLUSIONS}

In conclusion, we have investigated the field and temperature dependencies
of the magnetic anisotropy of $\mathrm{Ga}_{0.958}\mathrm{Mn}_{0.042}\mathrm{%
P}$ thin films synthesized by ion-implantation and pulsed laser melting
using measurements of the angular dependence of both ferromagnetic resonance
and SQUID magnetometry. The results of FMR and SQUID measurements including
coherent spin rotation and non-coherent spin switching can be understood
quantitatively using a relatively simple free energy model. Similar to $%
\mathrm{Ga}_{1-x}\mathrm{Mn}_{x}\mathrm{As}$ thin films, the magnetic
anisotropy is dominated by a strong out-of-plane uniaxial contribution.
Since the demagnetization field can only account for about one fourth of this
out-of-plane uniaxial anisotropy field, its most probable origin is in-plane
biaxial compressive strain, which is also the case for $\mathrm{Ga}_{1-x}%
\mathrm{Mn}_{x}\mathrm{As}$ thin films grown epitaxially on a GaAs
substrate. We also observe a cubic anisotropy contribution. However, the
sign of this cubic anisotropy term is opposite to the one commonly observed
for $\mathrm{Ga}_{1-x}\mathrm{Mn}_{x}\mathrm{As}$. While the latter finding could still be in agreement with Dietl's theory of hole-mediated ferromagnetism considering the significantly reduced hole concentration in $\mathrm{Ga}_{1-x}\mathrm{Mn}_{x}\mathrm{P}$ compared to $\mathrm{Ga}_{1-x}\mathrm{Mn}_{x}\mathrm{As}$, it remains to be demonstrated that this theory can indeed be applied to material systems exhibiting highly localized holes such as $\mathrm{Ga}_{1-x}\mathrm{Mn}_{x}\mathrm{P}$. Nevertheless, it is
an very interesting observation that in spite of the highly localized
character of the holes in $\mathrm{Ga}_{1-x}\mathrm{Mn}_{x}\mathrm{P}$, all
the magnetic properties (saturation magnetization, absolute values of the
anisotropy fields, Curie temperature) are similar to those typically
observed in $\mathrm{Ga}_{1-x}\mathrm{Mn}_{x}\mathrm{As}$. This constitutes an
important constraint for theories attempting to explain carrier-mediated
ferromagnetism in highly localized material systems. Finally, the
observation of symmetry-breaking in-plane uniaxial anisotropy components
similar to that seen in $\mathrm{Ga}_{1-x}\mathrm{Mn}_{x}\mathrm{As}$ $-$
where the origin still is under debate $-$ indicates an intrinsic origin
related to the hole-mediated ferromagnetic phase in III-Mn-V ferromagnetic
semiconductors. Moreover, because our samples were fabricated via II-PLM $-$ a form of liquid-phase epitaxy $-$ explanations brought forward that invoke vapor phase growth processes including effects of surface reconstruction can be excluded as the origin of this in-plane symmetry breaking.

\section*{ACKNOWLEDGMENTS}

We thank E. E. Haller for use of the ion implantation facilities and 
I. D. Sharp and J. W. Beeman for experimental assistance.
The work at the Walter Schottky Institut was supported by Deutsche
Forschungsgemeinschaft through SFB 631 and the Bavaria California Technology
Center; the work at Berkeley by the Director, Office of Science, Office of
Basic Energy Sciences, Division of Materials Sciences and Engineering, of
the U.S. Department of Energy under Contract No. DEAC02- 05CH11231 and
previous contracts.

\section*{APPENDIX A: ANISOTROPY ENERGY}

Magnetic anisotropy can e.g. be caused by dipole-dipole interaction, crystal
fields, and spin-orbit coupling. Furthermore, uniaxial and biaxial strain
can be an origin for magnetic anisotropies. While to date no comprehensive ab-initio
understanding of magnetic anisotropy has been established, magnetic anisotropy can be efficiently described mathematically with the help of
symmetry considerations. According to Chikazumi~\cite{Chi64}, the free energy of a uniaxial
anisotropy can be expressed by expanding it in a series of powers of $\sin
^{2}\vartheta $, 
\begin{equation}
F_{\mathrm{u}}=\tilde{K}_{\mathrm{u}1}\sin ^{2}\vartheta +\tilde{K}_{\mathrm{%
u}2}\sin ^{4}\vartheta +\ldots ~,  \label{eq:Euniax:chikazumi}
\end{equation}%
with the first and second order constants $\tilde{K}_{\mathrm{u}1}$ and $%
\tilde{K}_{\mathrm{u}2}$, respectively, and the angle $\vartheta $ between
the orientation of magnetization $\vec{m}=\frac{\vec{M}}{M}$ and the
anisotropy axis $\vec{u}$. This can be rewritten using $\sin
^{2}\vartheta =1-\cos ^{2}\vartheta $, so that 
\begin{eqnarray} \label{equ:uniax}
F_{\mathrm{u}} &=&\tilde{K}_{\mathrm{u}1}(1-\cos ^{2}\vartheta )+\tilde{K}_{%
\mathrm{u}2}(1-\cos ^{2}\vartheta )^{2}+\ldots \\
\nonumber &=&\left( \tilde{K}_{\mathrm{u}1}+\tilde{K}_{\mathrm{u}2}\right) +\left( -%
\tilde{K}_{\mathrm{u}1}-2\tilde{K}_{\mathrm{u}2}\right) \cos ^{2}\vartheta +%
\tilde{K}_{\mathrm{u}2}\cos ^{4}\vartheta +\ldots \\
\nonumber &=& \mathrm{const.}+K_{\mathrm{u1}}\cos ^{2}\vartheta +K_{\mathrm{%
u}2}\cos ^{4}\vartheta +\ldots , 
\end{eqnarray}%
with $K_{\mathrm{u1}}:=-\tilde{K}_{\mathrm{u}1}-2\tilde{K}_{\mathrm{u}2}$
and $K_{\mathrm{u}2}:=\tilde{K}_{\mathrm{u}2}$. Using $\vec{m}=\frac{\vec{M}%
}{M}={\left( 
\begin{array}{c}
\alpha _{x} \\ 
\alpha _{y} \\ 
\alpha _{z}%
\end{array}%
\right) }$ with the direction cosines of the cartesian axes $\alpha
_{x}=\sin \Theta \sin \Phi $, $\alpha _{y}=\cos \Theta $, and $\alpha
_{z}=\sin \Theta \cos \Phi $ from Fig.~\ref{fig:Koord}, the first order
uniaxial anisotropy contribution along $\vec{u}=\frac{1}{\sqrt{2}}{\left( 
\begin{array}{c}
0 \\ 
1 \\ 
1%
\end{array}%
\right) }$ in (\ref{FreeE}) for example is obtained via 
\begin{equation}
F_{\mathrm{u}}^{011}=K_{\mathrm{u}}^{011}\left( \vec{m}\vec{u}\right) ^{2}=%
\frac{1}{2}K_{\mathrm{u}}^{011}(\cos \Theta +\sin \Theta \cos \Phi )^{2}. \\
\end{equation}

The free energy for cubic magneto-crystalline anisotropy in cubic systems according to
Chikazumi~\cite{Chi64} is given by 
\begin{equation}
F_{\mathrm{c}}=K_{\mathrm{c}1}\left( \alpha _{x}^{2}\alpha _{y}^{2}+\alpha
_{y}^{2}\alpha _{z}^{2}+\alpha _{z}^{2}\alpha _{x}^{2}\right) +K_{\mathrm{c}%
2}\alpha _{x}^{2}\alpha _{y}^{2}\alpha _{z}^{2}+\ldots   \label{eq:Ecubic}
\end{equation}%
With the addition theorem for the direction cosines 
\begin{equation}
\alpha _{x}^{4}+\alpha _{y}^{4}+\alpha _{z}^{4}=1-2\left( \alpha
_{x}^{2}\alpha _{y}^{2}+\alpha _{y}^{2}\alpha _{z}^{2}+\alpha _{z}^{2}\alpha
_{x}^{2}\right) ,  \label{eq:dircos-hochvier}
\end{equation}%
(\ref{eq:Ecubic}) can be transferred to

\begin{equation}
F_{\mathrm{c}}= \mathrm{const.} - \frac{1}{2} K_{\mathrm{c}1} \left( \alpha _{x}^{4}+\alpha _{y}^{4}+\alpha _{z}^{4} \right) +K_{\mathrm{c}%
2}\alpha _{x}^{2}\alpha _{y}^{2}\alpha _{z}^{2}+\ldots .  \label{eq:Ecubic2}
\end{equation}

Therefore, the $\alpha _{i}^{4}$ terms in (\ref{eq:Ecubic2}) link the first-order cubic anisotropy and the second-order uniaxial anisotropy given in (\ref{equ:uniax}). Consequently, the latter formulation for first order cubic anisotropy is equivalent to a linear combination of second order uniaxial anisotropies along the cartesian axes, $F^i_{\mathrm{u2}}=K^i_{\mathrm{%
u}2} \alpha_i^4$, $i\in\left\{x,y,z\right\}$.

In (\ref{FreeE}) we accounted for the tetragonal crystal symmetry via distinguishing in-plane and out-of-plane cubic anisotropies.
Following the above discussion one could equivalently use a combination of a first order cubic anisotropy and a second order uniaxial anisotropy perpendicular to the film plane
\begin{eqnarray}
&-& \frac{1}{2} K^{\bot}_{\mathrm{c}1} \alpha _{x}^{4} - \frac{1}{2} K^{||}_{\mathrm{c}1} (\alpha _{y}^{4} + \alpha _{z}^{4}) = \\
\nonumber = &-& \frac{1}{2} K_{\mathrm{c}1} ( \alpha _{x}^{4} + \alpha _{y}^{4} + \alpha _{z}^{4}) + K^{100}_{\mathrm{%
u}2}\alpha_x^{4},
\end{eqnarray}%
with the first order cubic anisotropy constant $K_{\mathrm{c}1}=K^{||}_{\mathrm{c}1}$ and the second order uniaxial anisotropy constant perpendicular to the film plane $K^{100}_{\mathrm{%
u}2}=-\frac{1}{2} (K^{\bot}_{\mathrm{c}1}-K^{||}_{\mathrm{c}1})$.

Note also that because of
\begin{equation}
\alpha _{x}^{2}+\alpha _{y}^{2}+\alpha _{z}^{2}=1,  \label{eq:dircos-quadrat}
\end{equation}
only two of the three first order uniaxial anisotropy constants $K^{100}_{\mathrm{%
u1}}$, $K^{010}_{\mathrm{%
u1}}$, and $K^{001}_{\mathrm{%
u1}}$ are independent. A first
order uniaxial anisotropy can always be expressed by two other first order
uniaxial anisotropies perpendicular to each other. Analogously, the second-order uniaxial anisotropy constants also are not
independent because of Eq.\thinspace (\ref{eq:dircos-hochvier}).

\section*{APPENDIX B: DETERMINATION OF MAGNETIZATION}

Due to the Mn implantation profile, the magnetization can not be calculated as
usual via dividing the total magnetic moment $m_{\mathrm{tot}}$ measured e.g. via SQUID magnetometry by the sample volume. In a
first step we calculate the magnetic moment per \textit{substitutional} Mn atom via
\begin{equation}
m_{\mathrm{Mn}} = \frac{m_{\mathrm{tot}}}{D_{\mathrm{Mn, retained}} \cdot A \cdot f_{\mathrm{subst}}}, \label{eq:MperMn}
\end{equation}
where $D_{\mathrm {Mn, retained}}=7.3\times10^{15}$~cm$^{-2}$ is the Mn implantation dose retained after II-PLM, $A$ is
the sample area, and $f_{\mathrm subst}=0.7$ is the fraction of substitutional Mn atoms derived
via RBS and PIXE. To obtain an estimate for the magnetization of the sample in the region of highest Mn
concentration we multiply the average magnetic moment per Mn atom with the peak
substitutional Mn concentration $x$ and the density of Ga lattice sites
$\left[{\mathrm{Ga}}\right]=2.47 \times 10^{22}$~cm$^{-3}$
\begin{equation}
M = m_{\mathrm{Mn}} \cdot x \cdot \left[{\mathrm{Ga}}\right]. \label{eq:M}
\end{equation}

\end{document}